# Anti-mirror effect: A perfect lens brings a brighter feature

Yadong Xu [1], Shengwang Du [2], Lei Gao [1], and Huanyang Chen [1, *]

[1] School of Physical Science and Technology, Soochow University, Suzhou, Jiangsu 215006, China
[2] Department of Physics, The Hong Kong University of Science and Technology, Clear Water Bay, Kowloon, Hong Kong, China
* e-mail: kenyon@ust.hk

**Abstract:**

In this letter, we show that a perfect lens can be employed to make multiple objects appear like only one in the far field, leading to a new concept of illusion optics. Numerical simulations are performed to verify the functionalities for both passive and active objects. The conceptual device can be utilized to enhance the illumination brightness for both incoherent and coherent systems.

A perfect lens[1] made of a negative refractive index material can form ideal images beyond the diffraction limit. Recently, combined with transformation optics[2, 3, 4], it has been realized that many "magic" illusion effects can be obtained with extended perfect lens geometries[5, 6, 7]. In this work, we show that such a perfect lens can also act as an "anti-mirror" that makes multiple objects appear like only one in the far field. Our numerical simulation verifies that this "overlapped illusion optics" effect works for both passive and active objects (or sources). When applied to incoherent and coherent illumination systems, such as solid-state lighting, this technique can lead to dramatic enhancement in the illumination brightness and spatial-mode quality, as well as the heat-dissipation efficiencies.

It is well known that, as an object is placed in front of a plane mirror, a virtual image is formed on the other side. This image looks identical (except the opposite handedness) to the object viewed by an observer in front of the mirror. In other words, the mirror transforms a single object into two separate objects, as illustrated in Fig. 1a. One may ask an interesting question: Is this mirror effect invertible? Or, is there any way to make two objects look like one, which we shall call the "anti-mirror" effect? Our answer is yes! To illustrate the basic idea for simplicity, we consider here a two-dimensional (2D) case with transverse electric (TE) polarized waves. Two identical cylindrical perfect electric conductors (PECs) are placed on both sides of a perfect lens. The distance between them is *2d*, where *d* is the thickness of the perfect lens (see the schematic plot in Fig. 1b). Such a system displays the "anti-mirror" effect because the two PECs look like one to observers on both sides of the lens. More interestingly, the effect is also valid for active sources, as we will show later.

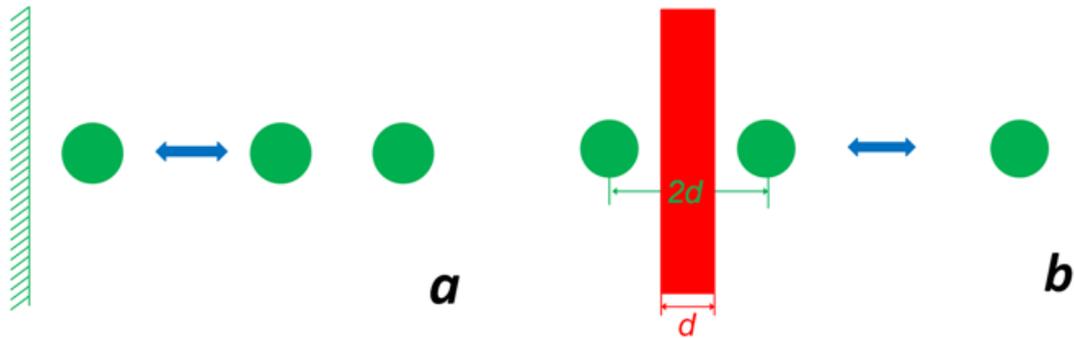

**Figure 1 Mirror and "anti-mirror" effect. a,** an object (green circle) in front of a plane mirror is equivalent to two identical objects (green circles). **b,** two identical PECs (green circles) on both sides of a perfect lens (red region) is equivalent to one PEC (green circle).

Illusion optics[7] and transformation optics[2, 3, 4] tell us that a perfect lens can be viewed as a transformation medium from a simple one-dimensional (1D) folded coordinate transformation[8]. The folded coordinate transformations can also bring a perfect lens with finite size. Here we use the folded coordinate transformations in Ref. [9] to illustrate the basic ideas. Figure 2a and 2b is the extension of Fig. 1b. In Fig. 2a, an elliptic cylindrical PEC is embedded in the restoring medium (blue regions) while an circular cylindrical PEC is located in another side of the perfect lens (red region) so that the positions and shapes of the two PECs (green regions) follow the image-forming principle of transformation optics[7]. As a result, the whole system in Fig. 2a looks like a bare circular cylindrical PEC (see in Fig. 2b) for the far-field observers. Such a phenomenon can even become more interesting. As we have known from illusion optics[7], an illusion device with an elliptic cylindrical PEC (green circle in Fig. 2c) embedded in the restoring medium (blue regions in Fig. 2c) looks like a bare circular cylindrical PEC (dashed green circle in Fig. 2c)[10]. Hence, we can replace the real cylindrical PEC in Fig. 2a with such an illusion device. Figure 2d gives a schematic plot showing that two objects look like one PEC (dashed green circle in Fig. 2d) to the far-field observers. *This effect has not been found in nature before.* In fact, we can understand it easily. The two illusion devices are close to each other, and their virtual illusion spaces[7] share a common region. Inside the shared region, the same PEC image is formed simultaneously by both illusion devices. Our numerical simulations show that this "overlapped illusion optics" (multiple objects look like one) works for both PECs and active sources.

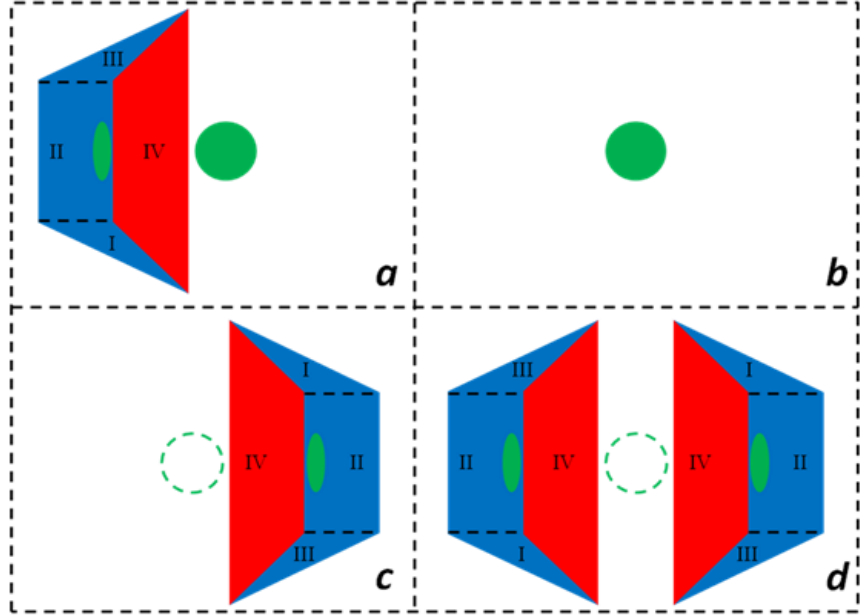

**Figure 2 Overlapped illusion optics. a,** the "anti-mirror" effect with finite size. **b,** a bare cylindrical PEC. **c,** an illusion device to form a virtual PEC image outside it. **d,** two illusion devices to form the same virtual PEC image, which shall be termed as the "overlapped illusion optics". The PECs are denoted by green color. The virtual PEC images are denoted by dash green circle. The perfect lenses are denoted by red color. The restoring media are denoted by blue color.

To demonstrate the above effect, we perform full-wave simulations using the COMSOL Multiphysics finite-element-based electromagnetics solver. We set the unit to be a wavelength. All the circular cylindrical PECs (both real and virtual) are located at the origin, whose radii are 0.4. The material parameters are related to the geometric shapes of the illusion devices. We use similar shapes to that in Ref. [9] ($p = -4/3$ for regions 'I', $p = 0$ for regions 'II', $p = 4/3$ for regions 'III', and $r = 1/3$). Regions 'I' are anisotropic materials with $\varepsilon_z = 3$, $\mu_{xx} = 17/3$, $\mu_{yy} = 3$, and $\mu_{xy} = \mu_{yx} = -4$. Regions 'II' are anisotropic materials with $\varepsilon_z = 3$, $\mu_{xx} = 1/3$, $\mu_{yy} = 3$, and $\mu_{xy} = \mu_{yx} = 0$. Regions 'III' are anisotropic materials with $\varepsilon_z = 3$, $\mu_{xx} = 17/3$, $\mu_{yy} = 3$, and $\mu_{xy} = \mu_{yx} = 4$. Regions 'IV' are perfect lenses with permittivity $\varepsilon = -1$ and permeability $\mu = -1$. The distances between the centers of the circular cylindrical PECs and the perfect lens interfaces are set to be 0.5, from which we can obtain the detailed shapes and positions of the elliptic cylindrical PECs. The TE waves is incident upward along the y-axis. Figure 3a shows the scattering patterns of the devices in Fig. 2a. The same rule applies in the rest parts of Fig. 3 and

Fig. 2. The identical far-field patterns in each part of Fig. 3 conform the above finding.

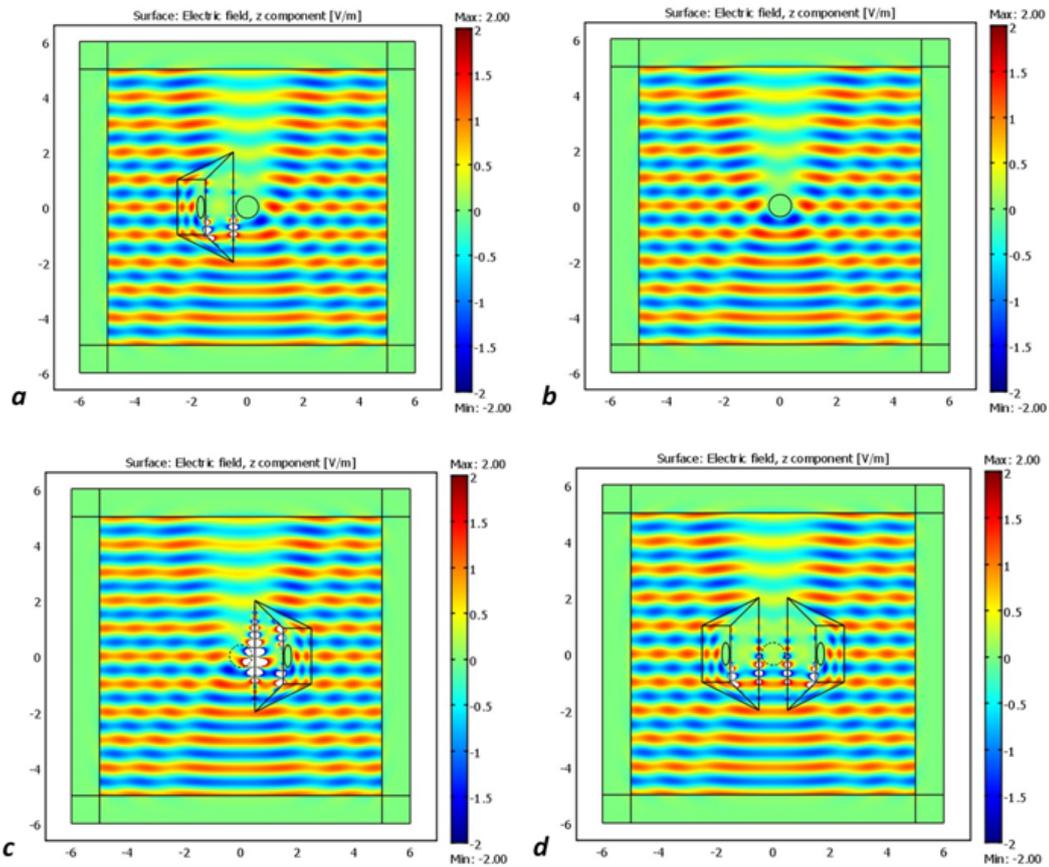

**Figure 3 Simulation results. a, b, c, and d** are the scattering patterns of the devices in Figs. 2a, 2b, 2c, and 2d.

The above overlapped illusion optics may provide solutions to modern solid-state illumination systems. Light-emitting diodes (LEDs) have been considered as the next generation lighting source because of their low operating voltage, small size, high energy-conversion efficiency and long lifetime. However, it is still a big challenge and a high cost to produce commercial single-LED bulbs to meet resident illumination requirement due to the heat dissipation and other manufacturing difficulties. To increase the illuminance level, a common solution is to package many LEDs inside a lamp. As a result, it is extremely difficult to generate spatial illumination uniformity for residence use. With the overlapped illusion optics proposed in this paper, this problem can be solved by overlapping the illusion images from all LEDs located physically at different positions - from an observer, it looks just like a single-LED source! Such a solid-state lighting device not only provides high illuminance level with spatial uniformity, but also dissipates heat efficiently.

Here we construct a model to simulate the proposed LED bulb. For comparison, in

Fig. 4a we show the intensity plot of the electric field ($|E_{z1}|^2$) of a single LED source simulated by a line source with $I_1 = 1A$ current and located at $\vec{r}_1 = (0, -0.5)$. The electric field $E_{z1}$ is[11] $E_{z1} = -\frac{1}{4}\frac{2\pi}{\lambda}\sqrt{\frac{\mu_0}{\varepsilon_0}}I_1 H_0^{(1)}(\frac{2\pi}{\lambda}|\vec{r}-\vec{r}_1|)$. When two LED sources (simulated by two incoherent sources) sit in parallel, as shown in Fig. 4b, the far-field light intensity is a sum of those of the two incoherent sources, with unavoidable spatial fluctuations. Here we set one source to be at $\vec{r}_1 = (0, -0.5)$ with a current $I_1 = 1A$ and the other at $\vec{r}_2 = (0, \frac{7}{6})$ with another current $I_2 = e^{i\varphi} A$, where $\varphi$ is a randomly different phase. To eliminate the intensity fluctuations at far field, we follow the proposed overlapped illusion optics and replace the two PECs in Fig. 2a (or Fig. 3a) with two line current sources. For aesthetic reasons, we rotate the device by 90 *deg* around the origin. The two line current sources are at the same positions of those in Fig. 4b. Region 'I' is anisotropic materials with $\varepsilon_z = 3$, $\mu_{xx} = 3$, $\mu_{yy} = 17/3$, and $\mu_{xy} = \mu_{yx} = 4$. Region 'II' is anisotropic materials with $\varepsilon_z = 3$, $\mu_{xx} = 3$, $\mu_{yy} = 1/3$, and $\mu_{xy} = \mu_{yx} = 0$. Region 'III' is anisotropic materials with $\varepsilon_z = 3$, $\mu_{xx} = 3$, $\mu_{yy} = 17/3$, and $\mu_{xy} = \mu_{yx} = -4$. Regions 'IV' are perfect lens with permittivity $\varepsilon = -1$ and permeability $\mu = -1$. Figure 4c shows the intensity distribution of the improved double-LED. Compared to Fig. 4b, it has the same level of light brightness but the device behaves like a single-LED bulb with perfect spatial quality. The result can be simply extended to many (>2)-LED bulbs.

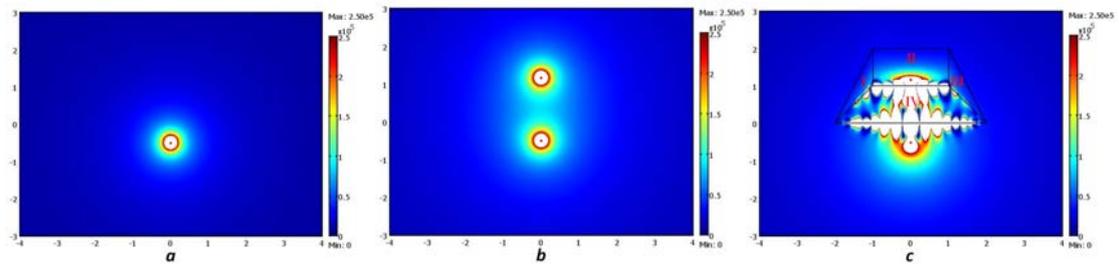

**Figure 4 An improved incoherent illumination system with overlapped illusion optics. a,** a single LED bulb: single line current source in vacuum. **b,** a double-LED bulb: double line current sources in vacuum. **c,** an improved double-LED bulb: double line current sources in the "anti-mirror" system. The intensity of the electric field are plotted.

When applied to coherent sources, our proposed method may be useful for developing high-power and high radiance coherent illumination sources with preserved spatial quality. In Fig. 5a, we simulate a single coherent source with a line current $I_1 = 1A$ at $\bar{r}_1 = (0, -0.5)$. In Fig. 5b, the spatial quality degrades when two coherent sources are aligned in parallel because of their interference. Figure 5c shows the simulation result of our anti-mirror effect (the same configuration as the LED bulb in Fig. 4c where the two LEDs are replaced with two coherent sources here). When the two sources are operated in the same optical frequency and phase, the light amplitude increases by a factor of 2 and thus the total power by a factor of 4! Such coherent system can be achieved using feed-back control with heterodyning detection. The increase of the output of the energy is not surprising because the overlapped illusion optics affects also the optical fields inside the two sources and increases the output conversion efficiency. This technique may have potential applications in beam-combining technique[12, 13] for developing high-power laser sources with preserved beam quality. Such effects cannot be obtained from the traditional beam-combining techniques.

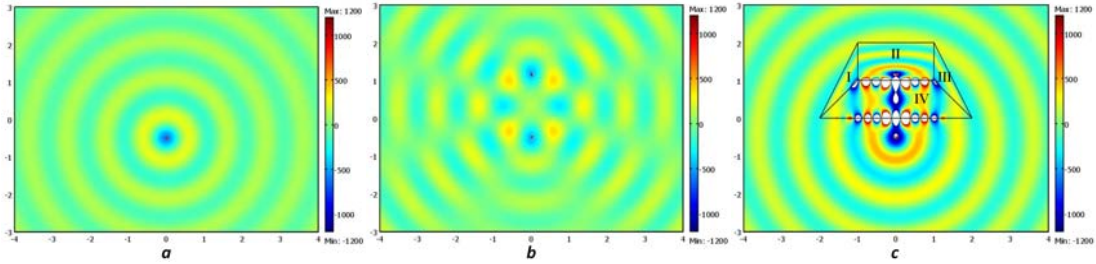

**Figure 5 Coherent sources with overlapped illusion optics (to motivate a new laser beam combining technique). a,** a single line current source in vacuum. **b,** two coherent line current sources in vacuum. **c,** a combined double-coherent source: two coherent line current sources in the "anti-mirror" system.

In summary, we have demonstrated the anti-mirror effect of the perfect lens. Transformation optics extends such an effect to make multiple objects look like one in the far field. Based on this concept, we proposed and numerically verified the overlapped illusion optics. When applied to incoherent illumination systems, we designed a many-LED bulb with brighter feature and much better spatial uniformity than a conventional one. Such a method may also be potentially applied in the beam-combining technique to generate high power coherent laser beams from multiple laser diodes (LDs) with preserved beam and spatial mode qualities. Therefore, the proposed anti-mirror effect and overlapped illusion optics may have wide applications.

## Acknoledgements

We thank Mr. Huihuo Zheng, Dr. Yun Lai, and Prof. C. T. Chan for their helpful discussions. This work was supported by the National Natural Science Foundation of China under Grant No. 11004147 and the Natural Science Foundation of Jiangsu Province under Grant No. BK2010211. S. Du was supported by the Hong Kong Research Grants Council.